\documentclass[lettersize,journal]{IEEEtran}
\usepackage{amsmath,amsfonts}
\usepackage{algorithmic}
\usepackage{algorithm}
\usepackage{array}
\usepackage[caption=false,font=normalsize,labelfont=sf,textfont=sf]{subfig}
\usepackage{textcomp}
\usepackage{stfloats}
\usepackage{url}
\usepackage{verbatim}
\usepackage{makecell}
\usepackage{graphicx}
\usepackage{multirow}
\usepackage{hyperref}
\usepackage{mathtools}
\usepackage{threeparttable} 
\usepackage{makecell}
\usepackage[table,dvipsnames]{xcolor}
\usepackage{nicematrix}
\usepackage{booktabs} 
\usepackage{graphicx} 

\usepackage[backend=biber,style=ieee,sorting=none]{biblatex}
\definecolor{inner}{RGB}{244,150,72}
\definecolor{indirect}{RGB}{110,136,230}
\definecolor{direct}{RGB}{187,240,129}

\usepackage{tikz}
\usetikzlibrary{decorations.pathreplacing,calc}

\begin{document}

\title{Double interior-point regularization for large-scale capacity expansion}

\author{Leonard Göke, Giovanni Sansavini
\thanks{Leonard Göke and Giovanni Sansavini are with the
Reliability and Risk Engineering Laboratory, Institute of Energy and Process
Engineering, ETH Zurich, 8092 Zurich, Switzerland.}
}



\maketitle

\begin{abstract}
Capacity expansion is a key tool for planning future energy systems. However, weather-dependent generation and long-duration storage result in problem sizes that exceed the computational limits of conventional interior-point solvers, making it impossible to plan renewable systems that are cost-efficient and reliable across a wide range of weather conditions. To tackle such large problems, this paper introduces the double interior-point regularization (DIP-set) for Benders Decomposition (BD), combining the advantages of traversing the interior of the solution space while remaining close to a reference solution. We benchmark the method on a power-sector problem and an energy-system problem, varying problem size and the level of foresight during operations. Results demonstrate that DIP-set outperforms competing regularizations in all test cases. The speed-up increases with size, reaching 30-50\% for the largest problems, which are the most critical for planning renewable systems and are too large for state-of-the-art methods. The key benefit of DIP-set is its ability to mitigate the sharp decrease in convergence as BD approaches the optimal solution.
\end{abstract}


\section{Introduction}

Mitigating climate change requires a fundamental transformation of power generation and consumption. Weather-dependent, and therefore variable and uncertain, generation from wind and solar energy replaces dispatchable thermal power plants. Consequently, the security of supply increasingly relies on both short- and long-duration energy storage \cite{Sepulveda2021}. At the same time, power consumption is shifting and rising, as direct or indirect electrification via e-fuels replaces fossil fuels in heating, transportation, and industry.

Capacity expansion models are essential tools for planning and guiding this transformation \cite{Hogan2002}; yet, modeling renewable and electrified systems is challenging. Capturing the weather-dependent variability of wind and solar generation requires high temporal and spatial resolution \cite{Goeke2021a}. In addition, short- and long-duration storage introduce complex interdependencies over extended time frames. Moreover, integrating large shares of renewables depends on transnational power grids, which expands the spatial scope of the models \cite{Victoria2020}. Finally, the need for electrification closely couples the power sector with heating, transport, and industry, necessitating an extended sectoral scope \cite{Brown2018}.

Advancing capacity expansion models to address these challenges is difficult. State-of-the-art multi-sector models with a continental scope and hourly resolution yield large linear optimization problems that approach the practical limits of established interior-point solvers, as memory requirements and computation time scale polynomial to exponential with model size. Increasing spatio-temporal detail, as suggested in the scientific literature, is not achievable with existing solution methods \cite{Frysztacki2021}. Moreover, current approaches hinder further methodological refinement, increasing the problem size; for instance, stochastic optimization to account for renewable uncertainty or robust optimization for planning a resilient transformation \cite{Plaga2023}.

To tackle these limitations, an emerging research stream applies Benders Decomposition (BD) for large-scale power and energy planning. BD is a solution algorithm that decomposes the original problem into a top-problem (TP) and one or more mutually independent sub-problems (SP) connected to the TP via so-called complicating variables \cite{Benders1962}. BD is particularly suited to two-stage stochastic problems, since each realization of uncertainty, or scenario, in the second stage can be an independent SP \cite{Slyke1969}. 

When BD is applied to capacity expansion in power systems, the complicating variables correspond to the capacities, which link investment decisions in the TP and the operational decisions in the SPs \cite{Conejo2006}. Since the SPs must be mutually independent, further decomposing them requires reformulations that add more complicating variables. \textcite{Jacobson2024} introduced temporal decomposition, adding complicating variables for storage levels; \textcite{Gruebler2025} benchmarked different degrees of spatial decomposition, adding complicating variables for energy flows between regions.

Standard BD suffers from oscillation between extreme solutions of the TP and, as a result, capacity expansion problems scale poorly with the number of complicating variables. Therefore, previous work suggested trust-region and level-set regularization to prevent oscillation by constraining the TP to solutions close to the current best solution \cite{Ruszczynski1986,Goeke2024}. This regularization improved convergence by up to two orders of magnitude on a linear problem, and subsequent studies confirmed the performance gains. However, the exact effect varied substantially across cases \cite{Wu2025,Sasanpour2025,Zampara2025}. 

On this basis, further studies applied an interior-point level-set that does not force the TP to be close to the current best solution, but yields interior, rather than corner-point, TP solutions \cite{Pecci2025, Zhang2025}. Moving through the interior benefits convergence, and, when tested on an integer problem, the interior-point level-set significantly improves performance compared to regularization approaches that enforce solutions close to the current best \cite{Gondzio2013}. At the same time, algorithmic refinements beyond regularization have not demonstrated comparable or consistent performance gains \cite{Goeke2024}.

In this paper, we introduce the interior-point regularization (DIP-set) that combines the strengths of trust-regions and interior-point level-set. The developed method traverses the interior of the solution space while remaining close to a reference solution. Combining both mechanisms is nontrivial, as it requires parametrizing two constraints such that both are simultaneously binding. We introduce two solutions for this problem: first, directly coupling both parameters, and second, reformulating the regularization to ensure indirectly that both constraints are binding simultaneously. Beyond BD, the method is also applicable to other decomposition-based or first-order solution methods.

After describing decomposition and BD in Section \ref{sec:bd}, Section \ref{sec:reg} presents the new regularization method. Section \ref{sec:cs} describes the two case studies. On this basis, Section \ref{sec:perf1} analyzes the performance of the new regularization method. Section \ref{sec:cl} concludes.

\section{Benders Decomposition for capacity expansion problems} \label{sec:bd}

This section describes how BD is applied to capacity expansion problems, including problem decomposition and the solution algorithm. The notations focus on a linear problem, but the following applies as long as all SPs are convex and non-convexity in the TP is limited to integer variables.

\subsection{Decomposition strategy}
 
The starting point of BD is the decomposition of a monolithic optimization problem into one TP and one or multiple SPs, resulting in the formulation given in Eqs. \ref{eq:1}. The variables $x$ and constraints in Eq. \ref{eq:1b} comprise the TP, while the variables $y$ and constraints in Eq. \ref{eq:1c} comprise the SPs. The variables $x$ connecting the TP and SPs are referred to as complicating variables:

\begin{subequations}
\begin{alignat}{2}
\min_{x,y} \; \; & c^\top x + d^\top y  \label{eq:1a} \\ 
s.t. \; \; & H x \leq a  \label{eq:1b}  \\   
& I x + J y \leq b \label{eq:1c} \\    
& x \in \mathbb{R}^{n}, y \in \mathbb{R}^{n} \label{eq:1d}
\end{alignat} \label{eq:1}  
\end{subequations}

\begin{figure}[!t]
\centering
\includegraphics[scale=1.0]{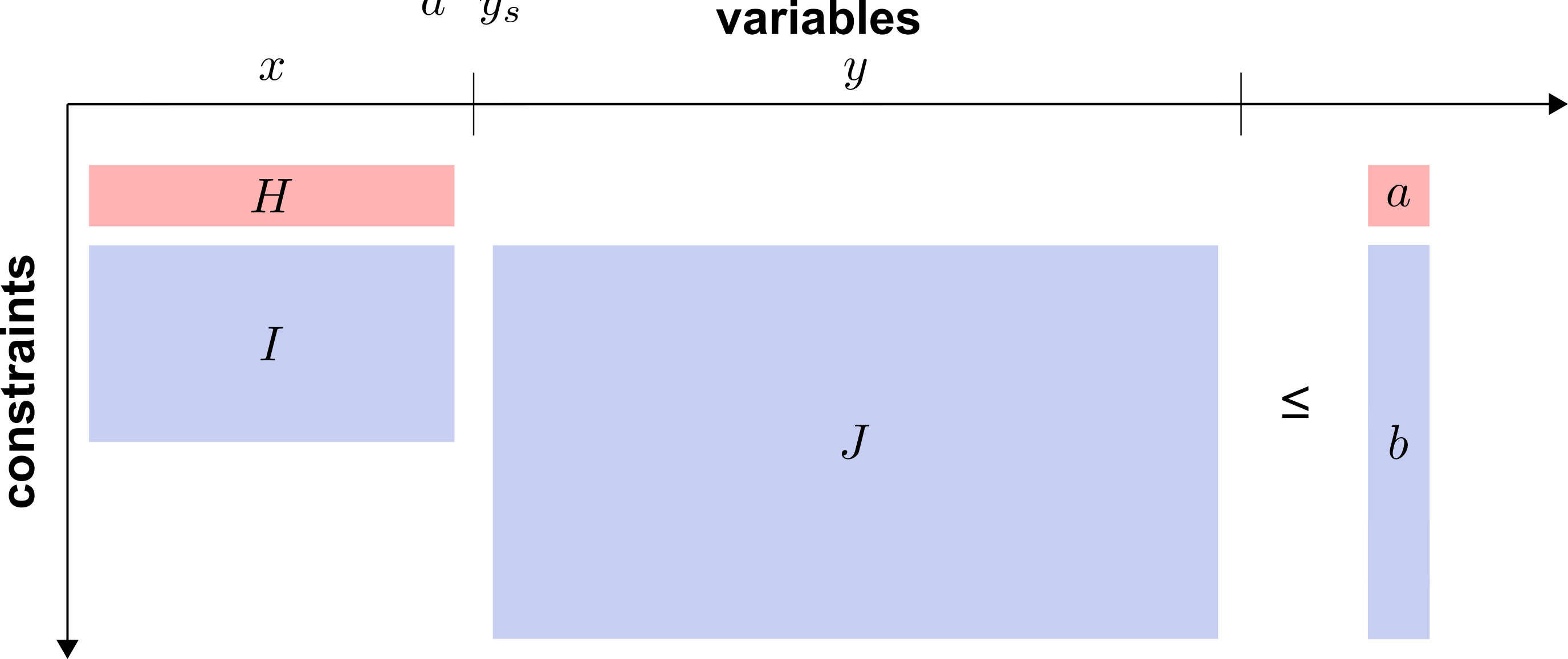}
\caption{Matrix structure of a decomposed problem with a single SP}
\label{fig:mono}
\end{figure}

\begin{figure}[!t]
\centering
\includegraphics[scale=1.0]{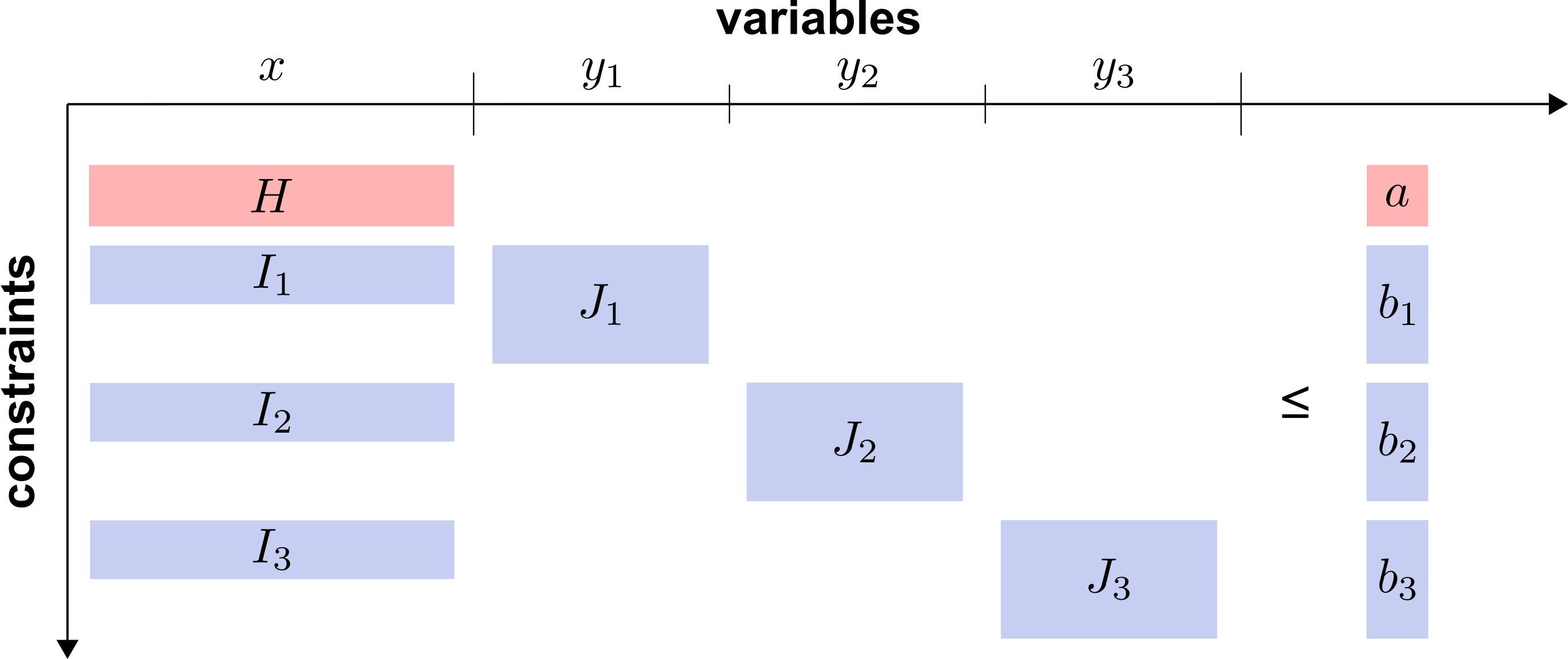}
\caption{Matrix structure of a decomposed problem with multiple SPs}
\label{fig:decom}
\end{figure}

Decomposing the problem is both a necessity and a key benefit. The memory requirements for solving a linear optimization problem depend on the size of all the matrices representing the problem, as shown in Figs. \ref{fig:mono} and \ref{fig:decom}. Memory requirements grow exponentially with the number of constraints and variables, which correspond to the columns and rows in the matrix, respectively. In capacity expansion problems, these matrices are extremely sparse, since most variables occur in only a few constraints. Accordingly, decomposing into smaller problems greatly reduces memory requirements.

In general, a sensible decomposition balances two objectives: first, keeping the number of complicating variables small to reduce the number of iterations required in the algorithm, and second, keeping the size of the TP and SPs small to reduce the computational effort per iteration. Moreover, decomposition into multiple SPs requires that SPs be mutually independent; i.e., the only variables that different SPs share are complicating variables. For example, Fig. \ref{fig:mono} shows a problem of the form in Eqs. \ref{eq:1} decomposed into one TP (red) and one SP (blue). If this SP has a block-diagonal structure, as shown in Fig. \ref{fig:decom}, the variables $y$ and constraints in Eq. \ref{eq:1c} can be partitioned into independent subsets, resulting in mutually independent SPs.

Capacity expansion problems are commonly decomposed into one TP representing capacity decisions, and one or more SPs for operation of the capacities. Accordingly, the variables $x$ are investment decisions and are constrained by Eq. \ref{eq:1b}, for instance, due to technical limits on installable capacity. The variables $y$ are operational decisions, for example, the electricity produced by a specific technology at a given time. The operational constraints of the SPs in Eq. \ref{eq:1c} can be divided into two categories. First, capacity constraints enforce that the operational decisions $y$ comply with the installed capacities $x$. These constraints connect the TP and the SPs and correspond to the rows associated with matrix $I$ in Figs. \ref{fig:mono} and \ref{fig:decom}. Second, the remaining constraints involve only operational decisions $y$, such as the energy balance, which ensures that supply meets demand at each time step.

Further decomposition of the SP depends on the problem structure. In a two-stage stochastic problem, the realizations of uncertainty, or scenarios, are mutually independent. As a result, the second stage can be decomposed into one SP per scenario. Further decomposition beyond scenarios depends on the storage formulation. Energy models track storage levels by computing the level $l_t$ at time step $t$ as the level in the previous time step $l_{t-1}$ plus the change in storage level $\delta_{t-1}$ in time step $t$. As a result, storage introduces time-coupling constraints that impose interdependencies between the different time steps of the operational problem. In light of these dependencies, there are two strategies to decompose the SP: 
\begin{enumerate}
\item \textbf{Limit dependencies:} Storage can be restricted to shift energy only within a limited time horizon, for instance, a day instead of the entire year. The cyclic constraint then forces the same storage level at the beginning and end of the day, rather than the same levels at the beginning and end of the year, preventing any dependencies that extend beyond the respective day. 

On the one hand, restricting storage to a limited horizon facilitates decomposition, but on the other hand, it limits the representation of storage technologies. When limiting the period to a day, the SP can be decomposed by day, resulting in small SPs. However, as a result, the modeled storage is limited to short-term storage, such as batteries, and cannot represent long-duration storage that shifts energy across seasons or years.

\item \textbf{Add levels as complicating variables:} The state of charge of storage systems, in short storage levels, can be added to the set of complicating variables $x$ \cite{Jacobson2024}. By deciding on levels in the TP, the SPs are again mutually independent, and decomposition into several SPs while considering long-duration storage is possible. This approach, however, increases the number of complicating variables and, therefore, the number of iterations in the later algorithm.
\end{enumerate}

For large capacity expansion problems, decomposition of the SP is necessary to solve the SPs in a reasonable time and fit them into memory. Therefore, decomposing two-stage stochastic problems by scenario is a no-regret choice. Further decomposition based on storage levels must strike a balance between increasing the number of complicating variables and decreasing the size of the SPs. Our recommended strategy is to include only long-duration storage levels as complicating variables, while restricting short-term storage to shifting energy within a limited time horizon \cite{Goeke2025}.

\subsection{Solution algorithm} \label{solAlgSection}

\begin{figure}[!t]
\centering
\includegraphics[scale=1.0]{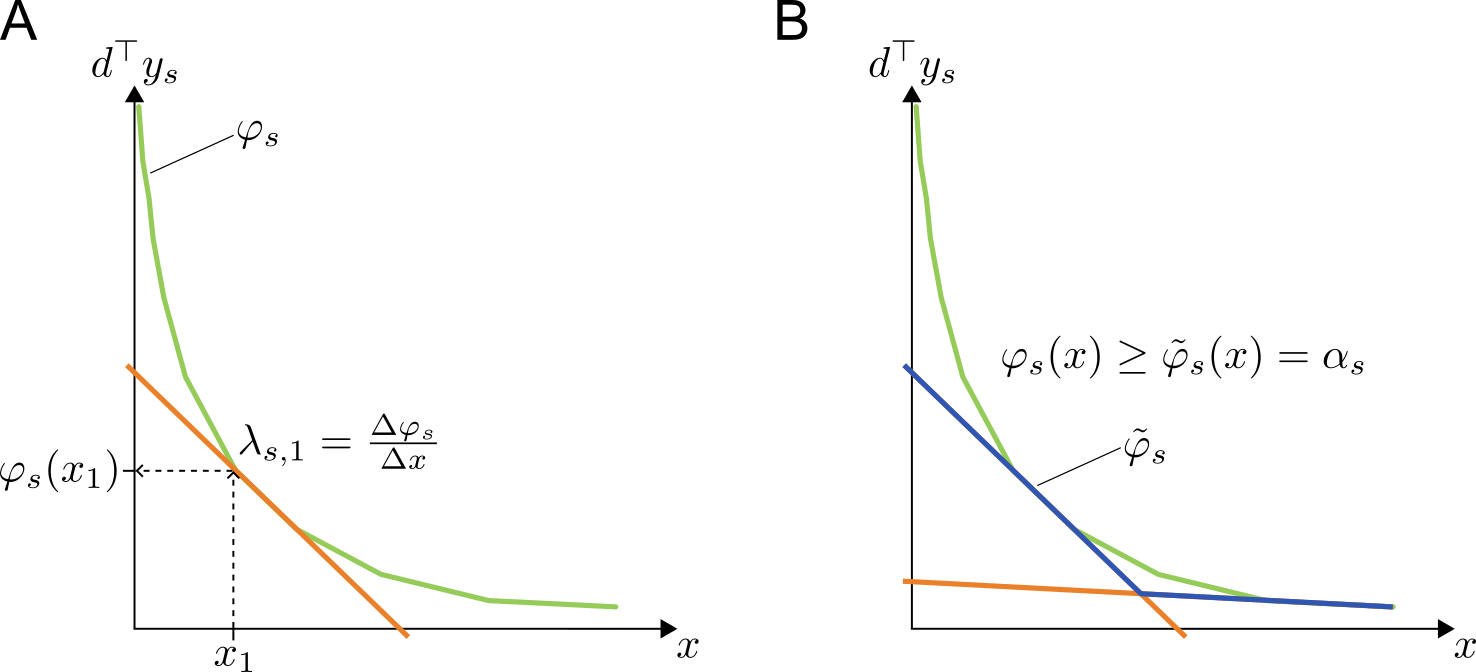}
\caption{The blue envelope of the organ cutting planes is a lower approximation to the green objective function of the convex SPs. 
Part (A) shows the addition of a cut in the first iteration; part (B) shows the envelope after adding one more cut in the second iteration.}
\label{fig:cut}
\end{figure}

To introduce the solution algorithm, we split the original problem in Eqs. \ref{eq:1} into one TP and several SPs, indexed over $s$. This notation is agnostic of whether the SP was decomposed by scenario, storage levels, or both. To solve the decomposed problem, Eq. \ref{eq:3} expresses the objective of each SP $s$ as the function $\varphi_s(x)$ of the complicating variables $x$:
\begin{equation}
\varphi_{s}(x) := \min\limits_{y_s \in \mathbb{R}^{n}}\{d^\top y_s \, | \, J_s y_s \leq b_s - I_s x\} \label{eq:3}
\end{equation}
Adding the function $\varphi_s(x)$ to the objective of the TP in Eq. \ref{eq:2} gives a decomposed formulation for the original problem:
\begin{subequations}
\begin{alignat}{2}
\min_x \; \; & c^\top x + \sum_{s \in S} \varphi_{s}(x) \label{eq:2a} \\
s.t. \; \; & H x \leq a \label{eq:2b} \\  
& x \in \mathbb{R}^{n} \label{eq:2c}
\end{alignat} \label{eq:2} 
\end{subequations}
BD then replaces the unknown function $\varphi_{s}(x)$ with its lower approximation $\tilde{\varphi}_{s}(x) \leq \varphi_{s}(x)$. Since the SPs are convex, the approximation can be built from linear cutting planes enveloping the original function. The cutting planes are constructed from the objective value $z_{s,i} = \varphi_{s}(x_i)$ and dual variables $\lambda_{s,i}$ when solving the SPs for specific values of the complicating variables $x_i$ in iteration $i$. On a one-dimensional example, Fig. \ref{fig:cut} illustrates how the orange cutting planes build the blue approximation of the true function in green.

The envelope of cutting planes can be added to the TP by replacing $\varphi_{s}(x)$ in Eq. \ref{eq:2a} of the TP with the variable $\alpha_{s}$ and adding constraints that enforce all cutting planes as lower bounds of $\alpha_{s}$, as described in Eq. \ref{eq:4}.
\begin{equation}
\begin{split}
\tilde{\varphi}_{s}(x)
&= \max\limits_{i \in N}\{z_{s,i} + \lambda_{s,i}^\top (x - x_i)\} \\
&\Leftrightarrow \alpha_{s} \geq z_{s,i} + \lambda_{s,i}^\top (x - x_i),
\quad \forall i \in N
\end{split}
\label{eq:4}
\end{equation}

In capacity expansion models, the function $\varphi_{s}(x)$ gives the operational costs for given capacities and storage levels. To ensure the SPs are feasible for any $x$, the SPs include slack variables, for instance, a loss-of-load variable that guarantees the energy balance can always be met, albeit at extremely high cost. As a result, implementations of BD in energy planning do not require feasibility cuts to prevent infeasible SPs.

The standard Benders algorithm in Alg. \ref{alg:alg1} repeatedly solves the TP, including Benders cuts, to obtain a candidate solution $x_i$ for the complicating variables and then solves each SP for this candidate. Since all SPs are mutually independent, solving all SPs can leverage parallelization and distributed computing.

Thanks to the lower approximation provided by Benders cuts, the TP objective value provides a lower bound $\zeta^{low}_i$ on the optimal solution. The objective value of the TP, excluding the cuts, and summed with the objective value of the SPs gives the true objective value for the candidate solution $x_i$. The smallest true objective value across iterations provides an upper bound $\zeta^{up}_i$ on the optimal solution.
{\linespread{1.1}\selectfont
\begin{algorithm}[H]
\caption{Standard Benders algorithm}
\begin{algorithmic}
\STATE 
\STATE \textbf{while} $1-\zeta^{low}_i/\zeta^{up}_i>\epsilon$ 
\STATE \hspace{0.5cm} $x_i \gets \text{Solve TP in Eq. \ref{eq:2} with Benders cuts in Eq. \ref{eq:4}}$   
\STATE \hspace{0.5cm} $z_{s,i}, \lambda_{s,i} \gets \text{Solve SP in Eq. \ref{eq:3} for } x_i$
\STATE \hspace{0.5cm} $\zeta^{low}_i \gets c^\top x_i + \displaystyle \sum_{s \in S} \tilde{\varphi}_{s}(x_i)$
\STATE \hspace{0.5cm} $\zeta^{up}_i \gets \min\{\zeta^{up}_i, c^\top x_i + \displaystyle \sum_{s \in S} \varphi_{s}(x_i)\}$
\STATE \hspace{0.5cm} Update $\tilde{\varphi}_{s}$ with $x_i, z_{s,i}$ and $\lambda_{s,i}$ 
\STATE \textbf{end while}
\end{algorithmic} \label{alg:alg1}
\end{algorithm}
}
At the end of an iteration, Benders cuts based on the objective value $z_{s,i}$ and dual variables $\lambda_{s,i}$ from the solution of the SPs in the current iteration are added to the TP. As a result of these additions, the TP yields a new solution in the following iteration, based on a more precise approximation of the SPs, guaranteeing the algorithm will eventually converge to an optimal solution. However, in practice, a target optimality gap $\epsilon$ is used as a stopping criterion.

In the Benders cuts, there is a fundamental difference between the complicating variables for capacities, which connect the TP and SPs, and those for storage levels, which decompose operations into independent SPs. Since adding capacity cannot increase the operational costs of the SPs, the dual variables of the capacity variables are always non-positive, but other complicating variables can have positive and negative dual values. For instance, a complicating variable for the storage level at time-step $t$ yields a non-positive value for SPs that start at time-step $t$, since increasing the initial level offers additional energy to the SP. However, the same complicating variable yields a non-negative value for SPs that end at time-step $t$, since increasing the final level withdraws energy from the SP. As a result, complicating variables for storage levels and capacity affect convergence differently.

\section{Regularization of Benders Decomposition} \label{sec:reg}

The standard Benders algorithm is guaranteed to converge, but the convergence rate deteriorates with the number of complicating variables; as a result, large capacity expansion problems converge extremely slowly. Regularization methods greatly improve performance by constraining the candidate solution $x_i$ in the TP to be close to a reference $x_r$, thereby preventing oscillation of the algorithm. 

This section discusses regularization methods and their general implementation. Afterward, Subsection \ref{exStab} describes specific existing regularization methods and Subsection \ref{newStab} introduces the double interior-point regularization (DIP-set).  

Alg. \ref{alg:alg2} shows how regularization generally changes the BD algorithm, regardless of the specific method. Before the iterations begin, the regularized algorithm initializes the reference solution $x_r$, for example, by solving a heuristic version of the original problem using time-series reduction. During iteration, a regularized version of the TP is solved to propose a candidate solution, $x_i$, that is close to the reference solution $x_r$. If the candidate solution improves the upper bound, it becomes the new incumbent solution, and $x_r$ is updated accordingly. Finally, the original TP without regularization is solved to obtain an accurate lower bound.

{\linespread{1.1}\selectfont
\begin{algorithm}
\caption{Regularized Benders algorithm}\label{alg:alg2}
\begin{algorithmic}
\STATE 
\STATE Initialize $x_r$
\STATE \textbf{while} $1-\zeta^{low}_i/\zeta^{up}_i>\epsilon$ 
\STATE \hspace{0.5cm} $x_i \gets \text{Solve regularized TP with Benders cuts in Eq. \ref{eq:4}}$   
\STATE \hspace{0.5cm} $z_{s,i}, \lambda_{s,i} \gets \text{Solve SP in Eq. \ref{eq:3} for } x_i$
\STATE \hspace{0.5cm} $\zeta^{low}_i \gets \text{Solve TP in Eq. \ref{eq:2} with Benders cuts in Eq. \ref{eq:4}}$   
\STATE \hspace{0.5cm} \textbf{if} $\zeta^{up}_i > c^\top x_i + \displaystyle \sum_{s \in S} \varphi_{s}(x_i)$
\STATE \hspace{1.0cm} $x_r \gets x_i$
\STATE \hspace{1.0cm} $\zeta^{up}_i \gets c^\top x_i + \displaystyle \sum_{s \in S} \varphi_{s}(x_i)$
\STATE \hspace{0.5cm} \textbf{end if}
\STATE \hspace{0.5cm} Update $\tilde{\varphi}_{s}$ with $x_i, z_{s,i}$ and $\lambda_{s,i}$ 
\STATE \textbf{end while}
\end{algorithmic}
\end{algorithm}
}

\begin{figure}[!t]
\centering
\includegraphics[scale=1.0]{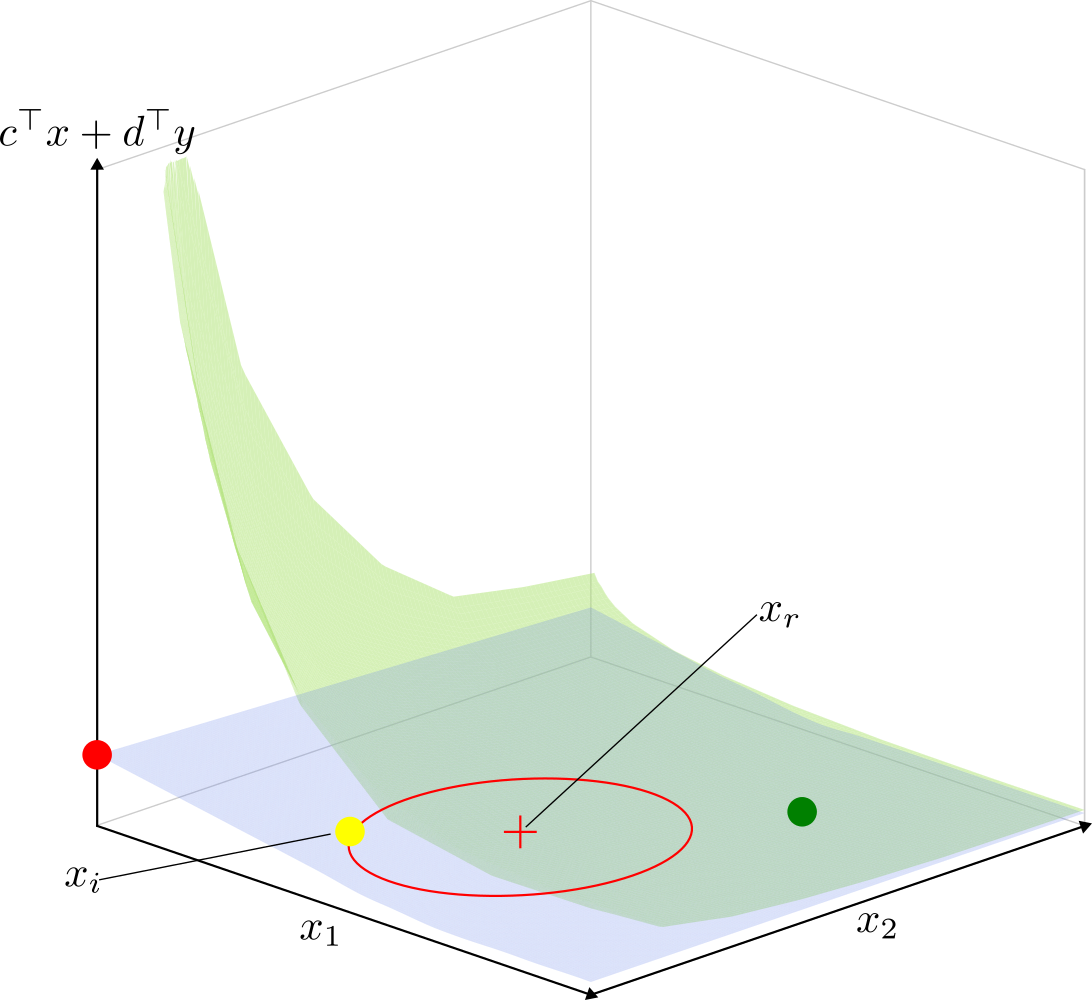}
\caption{Regularization restricts the current iteration to a region around the current best solution $x_r$. It prevents moving to extreme points based on the current blue approximation of the entire problem, where the difference to the true problem in green is large.}
\label{stag}
\end{figure}

Regularized BD remains guaranteed to converge to an optimal solution. To illustrate, the green plane in Fig. \ref{stag} shows the objective value as a function of two complicating variables; the blue plane shows the current approximation. Due to regularization, the TP is restricted to the red circle, and therefore yields the yellow candidate solution instead of the red one. Only if the candidate solution is closer to the unknown green optimum than the reference solution, the candidate will improve the current best and become the new reference solution. As a result, the reference solution monotonically approaches the optimum. Convergence of the original algorithm ensures that within the red circle, regularized BD finds a solution that improves the reference solution.

All regularization methods use the Euclidean ($l_2$) norm as a distance measure. Although other norms are theoretically possible, the $l_\infty$-norm performs poorly in practice, and all remaining norms are difficult to implement \cite{Goeke2024}. Note that the $l_2$-norm results in a quadratic formulation for integer or continuous variables, whereas for binary variables it can be represented using linear constraints \cite{Hill2018}.

Moreover, all regularization methods rely on at least one dynamic parameter that controls the acceptable distance between the new candidate solution $x_i$ and the reference $x_r$. The literature lacks generalizable insights into how different strategies for setting the dynamic parameter affect performance, which introduces a risk of overfitting and cherry-picking. Accordingly, this paper uses standard methods and discusses the parameter strategy only when relevant to the case studies presented in Section \ref{sec:cs}.

Other refinements beyond regularization, such as valid inequalities or inexact cuts, have already been investigated in previous publications, or, in the case of advanced cut generation, initial tests showed very little promise \cite{Goeke2024, Sherali2013}.

\subsection{Existing regularization methods} \label{exStab}

As the first specific regularization, we introduce the trust-region in Eq. \ref{eq:5} \cite{Goeke2024}. In this case, a quadratic constraint directly limits the distance between $x$ and the reference solution $x_r$, as illustrated in Fig. \ref{stag}. The dynamic parameter $r_i$ is the upper limit on the distance, or, in geometrical terms, the radius of a hypersphere constraining the TP.

\begin{subequations}
\begin{alignat}{2}
\min_x \; \; & c^\top x + \sum_{s \in S} \tilde{\varphi}_{s}(x) \\
s.t. \; \; & ||x - x_r||_2  \; \; \leq r_i
\end{alignat} \label{eq:5}
\end{subequations}

Next, the proximal regularization in Eq. \ref{eq:51} corresponds to a Lagrangian relaxation of the trust-region, and, thus, the method penalizes the distance from the reference in the objective function. The dynamic parameter is the Lagrange multiplier $\tau_i$ \cite{Lemarechal1981}.

\begin{subequations}
\begin{alignat}{2}
\min_x \; \; & c^\top x + \sum_{s \in S} \tilde{\varphi}_{s}(x)  + \tau_i ||x - x_r||_2 
\end{alignat} \label{eq:51}
\end{subequations}

Third, the level-set regularization in Eq. \ref{eq:6} directly minimizes the distance between the complicating variables and the reference solution; an additional constraint limits the original objective to $\lambda_i$, the dynamic parameter \cite{Ruszczynski1986}. 
\begin{subequations}
\begin{alignat}{2}
\min_x \; \; & ||x - x_r||_2 \\
s.t. \; \; & c^\top x + \sum_{s \in S} \tilde{\varphi}_{s}(x) \leq \lambda_i \label{eq:6b}
\end{alignat} \label{eq:6}
\end{subequations}

To prevent the regularized TP from resulting in the reference solution again, $\lambda_i$ must be below the objective value at the reference solution, namely, the current upper bound $\zeta^{up}_i$. Vice versa, $\lambda_i$ must be above the lower bound $\zeta^{low}_i$ to prevent an infeasible regularized problem. Therefore, the common strategy is to set $\lambda_i$ to a weighted average between $\zeta^{low}_i$ and $\zeta^{up}_i$ using the weight $\beta$, as in Eq. \ref{eq:7}. If $\lambda_i$ results in an infeasible problem, the lower bound $\zeta^{low}_i$ is increased to the current $\lambda_i$ and a new value is computed.

\begin{equation}
\lambda_i = \beta\zeta^{low}_i + (1-\beta)\zeta^{up}_i \label{eq:7}  
\end{equation}

The interior-point level-set regularization in Eq. \ref{eq:8} results in a solution in the interior of the solution space \cite{Gondzio2013}. To this end, the variant replaces the distance minimization in the level-set regularization with an empty objective function, resulting in a feasibility problem. Accordingly, this regularization no longer utilizes the reference solution $x_r$. Instead, the interior-point level-set solely relies on the lower and upper bounds that determine $\lambda_i$ to regularize the TP and accelerate convergence towards the optimal solution.
\begin{subequations}
\begin{alignat}{2}
\min_x \; \; & 0 \\
s.t. \; \; & c^\top x + \sum_{s \in S} \tilde{\varphi}_{s}(x) \leq \lambda_i
\end{alignat} \label{eq:8}
\end{subequations}

\subsection{Double interior-point regularization (DIP-set)} \label{newStab}

The introduced double interior-point regularization (DIP-set) in Eq. \ref{eq:81} builds the interior-point level-set regularization and returns a solution in the interior of the solution space. However, DIP-set still utilizes information from the reference solution $x_r$. For this purpose, it adds the quadratic constraint from the trust-region regularization to the problem. By traversing the interior of the solution space while remaining close to the reference solution, DIP-set combines the strengths of trust-region and interior-point level-set regularization.

\begin{subequations}
\begin{alignat}{2}
\min_x \; \; & 0 \\
s.t. \; \; & c^\top x + \sum_{s \in S} \tilde{\varphi}_{s}(x) \leq \lambda_i \\
& ||x - x_r||_2  \; \; \leq r_i
\end{alignat}  \label{eq:81}
\end{subequations}

Since DIP-set combines two regularizations, it also has two dynamic parameters: the upper limit on the original objective value $\lambda_i$, and the trust-region radius $r_i$. On the one hand, the choice of the dynamic parameters should not make the TP infeasible; on the other hand, both regularizing constraints must be binding for the method to be effective. A direct and an indirect implementation of DIP-set both use the common strategy for the upper limit $\lambda_i$, described in Eq. \ref{eq:7}, but differ in how they handle the second parameter.

In the direct implementation, the dynamic parameter for the trust-region is directly coupled to the optimality gap. The radius $r_i$ is computed from an initial radius $r_0$, the sum of all complicating variables, multiplied by a percentage from an interval from 0.5 to 10\%, as described in Eq. \ref{eq:9}. An interpolation $\mathcal{I}$ based on the current optimality gap yields the specific value within this interval; thus, the radius decreases as the algorithm progresses. 
\begin{equation}
r_i = r_0 \cdot \mathcal{I}(1-\frac{\zeta^{low}_i}{\zeta^{up}_i};\, 0.1, 0.05) \label{eq:9}
\end{equation}

The indirect implementation leverages the fact that the DIP-set feasibility problem in Eq. \ref{eq:8} corresponds to a suboptimal termination of the original level-set in Eq. \ref{eq:6}. Both regularizations yield the same result if the level-set terminates once the distance between $x$ and $x_r$ is below the threshold $r_i$ instead of solving for the smallest distance possible. Therefore, instead of setting $r_i$ and solving the TP in Eq. \ref{eq:8}, the DIP-set regularization can also be implemented by using a loose convergence criterion and solving the level-set TP in Eq. \ref{eq:6}. This implementation can only control the radius indirectly, but facilitates ensuring that both regularizations are binding while maintaining feasibility.

\section{Case studies} \label{sec:cs}

We test BD and the DIP-set regularization across various configurations of two capacity expansion problems to ensure robustness. Both problems are linear, but the results directly transfer to integer problems since previous research has shown that BD solves capacity expansion problems best by first solving the linear relaxation, which accounts for the majority of the computation time \cite{Pecci2025}.

The two capacity expansion problems plan a net-zero system for a spatially resolved European continent by determining capacities for generating, converting, storing, and transporting energy carriers. The first capacity expansion covers only the power sector, while the second, more elaborate problem covers the entire energy system.

The power-sector and energy-system problem, along with the corresponding BD algorithm, are embedded in the open-source modeling framework AnyMOD.jl \cite{Goeke2020a}. For the general mathematical formulation and in-depth description of both problems, see \textcite{Goeke2024} and \textcite{Goeke2025}.

Both capacity expansion problems are two-stage stochastic problems. Capacity decisions in the first stage are subject to operational uncertainty in the second stage. In the case studies, the second-stage scenarios correspond to different weather years, but the presented methods are transferable to any operational uncertainty. 

\begin{figure}[!t]
\centering
\includegraphics[scale=1.0]{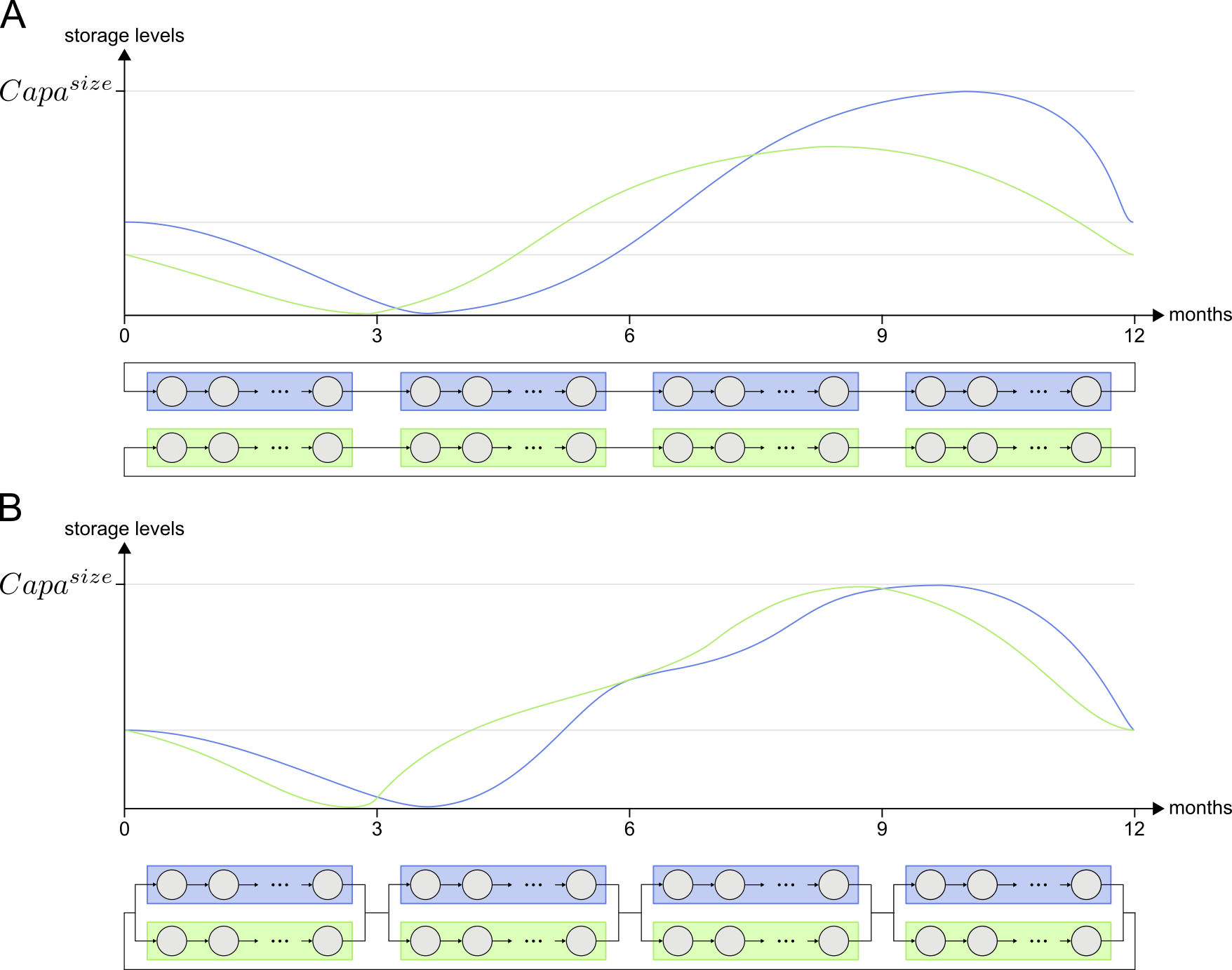}
\caption{Exemplary storage levels under perfect foresight (A) and limited foresight (B) in the case of two weather years, each decomposed into four SPs with three months.}
\label{figFrs}
\end{figure}

The benchmarks vary in the level of foresight about future conditions during system operation, resulting in different time-coupling constraints for energy storage and, thus, enabling tests of BD with different decomposition strategies. The two approaches are:

\begin{enumerate}
\item \textbf{Perfect Foresight}: Under perfect foresight in part (A) of Fig. \ref{figFrs}, future conditions are perfectly known in advance. Perfect foresight implies that operational decisions only anticipate the future conditions of the same year. For example, the first period of the blue year in Fig. \ref{figFrs} only depends on the other blue periods.

Accordingly, the time-coupling constraints for storage, indicated by the arrows, connect only periods within the same year. A cyclic condition enforces the same level at the beginning and end of each year to close the energy balance for storage. Storage levels across distinct years are independent, except for a shared upper limit imposed by the storage's energy capacity $Capa^{size}$, determined in the first stage. The perfect foresight approach is the standard and has been used in all previous studies on BD for capacity expansion.

\item \textbf{Limited Foresight}: Limited foresight in part (B) of Fig. \ref{figFrs} assumes that conditions are only perfectly known within each period; conditions beyond the period are stochastically independent, i.e., the likelihood of the blue or green year in the second period does not depend on which year occurred in the first period, and so on. 

Limited foresight implies that storage levels are uniform at the beginning and end of periods across different years, as shown in the storage curves in part (B). This constraint reflects that, in line with our assumptions, system operation cannot anticipate specific future conditions. The limited foresight approach was introduced in \textcite{Goeke2025}.

\end{enumerate}

With perfect foresight, each SP can cover up to one full year, or the SPs can further decompose the year into shorter periods at the expense of adding complicating storage levels to the TP. Under limited foresight, uniform storage levels impose dependencies across years. As a result, the SPs must decompose the year into shorter periods that correspond to the level of foresight. Again, complicating storage levels in the TP connect the periods, but fewer variables are required than in the perfect foresight case, since the uniform levels at the beginning and end of periods reduce the degree of freedom.

To reduce the computational impact of storage modeling, both problems apply the described approaches only for long-term storage. Technologies that can only shift energy over short time frames, such as batteries, instead have a separate cyclic condition for each period to reduce model complexity.

\section{Performance of the DIP-set regularization} \label{sec:perf1}

This section analyzes how the DIP-set regularization performs compared to the alternatives from the literature. Although the ultimate criterion for BD performance is time to convergence, the analysis focuses on the number of iterations required. The time to convergence equals the number of iterations multiplied by the average time per iteration. However, the time per iteration depends on the performance of the solver we use for the TP and SPs, i.e., Gurobi's Barrier algorithm \cite{gurobi}. This performance is non-deterministic and, in our case, also depends on the hardware allocated to a specific computation by the deployed computing cluster. As a result, computation times are biased, and we instead benchmark the regularization based on the number of iterations. Beforehand, we confirmed that there was no correlation between time per iteration and the regularization method. We also ensured all runs in the benchmark converged to the correct objective value.

All runs perform a heuristic presolve to obtain a starting solution and initialize the regularization. This presolve solves a deterministic version of the problem, using only the most probable scenario and reducing the time series to 672 hours.

Furthermore, we limit the trust-region constraint to the complicating variables for capacities, since including complicating storage variables consistently deteriorated performance. That complicating storage variables, in contrast to capacity variables, do not benefit from a trust-region constraint is plausible, given that storage variables can have positive and negative duals in the Benders cuts, whereas capacity variables have strictly non-negative duals, as discussed in Subsection \ref{solAlgSection}.

\subsection{Pre-selection of parameter strategy} \label{sec:perf2}

The performance of regularization depends on the strategy for choosing its dynamic parameters, but the impact is seemingly chaotic, and existing research offers no clear criteria for setting parameters. Consequently, any benchmark is at risk of cherry-picking a parameter strategy that works well for a single problem but is not transferable \cite{Goeke2024}.

To mitigate this risk, we pursue a ``one-size-fits-all'' approach. First, we solve a simplified problem across a wide range of parameter strategies, then select the most promising candidates for extensive testing with both case studys. The approach also reflects that, in practice, testing a large number of parameter strategies defeats the purpose of reducing computational effort.

As a simplified problem, we use the power-sector problem for all 33 regions, but with only 672 instead of 8,760 hours. Each SP covers one scenario, or 12 months; i.e., there is no further decomposition or limited foresight, and thus no complicating storage variables. Compared to the same problem with all 8,760 hours, the simplification underestimates system costs by 12\%. 

The initial benchmark compares the methods and parameter strategies listed in Tab. \ref{tab:tab1}, all of which build on the level-set method and its standard parameter strategy. For the direct implementation of the DIP-set regularization, we test linear, exponential, and logarithmic interpolation to set the radius based on the current optimality gap. For the indirect implementation, we test convergence tolerances of 0.1, 0.5, and 1.0 by setting the \textit{BarConvTol} option in Gurobi's Barrier algorithm to achieve earlier termination. The standard level-set method uses the default value of 1e-8.

We exclude trust-region and proximal regularization because they are theoretically equivalent to level-set methods, and we run BD until its optimality gap is below 0.1\%.

\begin{table}
\begin{center}
\caption{Tested configurations for considered regularization methods}
\label{tab:tab1}
\renewcommand{\arraystretch}{1.2}
\scriptsize
\begin{tabular}{| c | c | c |}
\hline
\multicolumn{2}{|c|}{\textbf{regularization method}} & \textbf{tested level-set parameter} \\
\Xhline{0.8pt}
\multicolumn{2}{|c|}{level-set}  & \multirow{3}{*}{$\beta \in \{0.125,0.25,...,0.875\}$} \\
\cline{1-2} \multicolumn{2}{|c|}{interior-point level-set}  &  \\
\cline{1-2} \multirow{2}{*}{DIP-set} & indirect  &   \\
\cline{2-3}  & direct  & $\beta \in \{0.25,0.5,0.75\}$   \\
\hline
\end{tabular}
\end{center}
\end{table}

Tab. \ref{tab:tab2} presents the results of the initial benchmark. Regardless of the level-set parameter $\beta$, the standard level-set method performs poorly. Interior-point level-set regularization performs best with $\beta$ set to 0.375.
Both results confirm \textcite{Pecci2025}, who also find that the interior-point level-set is superior and performs best with $\beta$ set to 0.5.

The indirect implementation of DIP-set regularization outperforms the direct implementation overall, which is plausible given that it can enforce the additional trust-region constraint more effectively. The correlation between $\beta$ and the number of iterations is similar to the other regularizations, and the best performance is achieved with $\beta$ set to 0.25 and a termination criterion of 0.5.

\begin{table}[ht]
\centering
\caption{Number of iterations for power-sector problem to converge with 33 regions, perfect foresight, 12 months per SP reduced to 672 hours.}
\label{tab:tab2}

\setlength{\extrarowheight}{2pt}

\begin{tikzpicture}
\node (tbl) {
\renewcommand{\arraystretch}{1.2}
\scriptsize
\begin{tabular}{|
c | c | c
    !{\vrule width 0.8pt} >{\centering\arraybackslash}m{0.5cm}
    | >{\centering\arraybackslash}m{0.5cm}
    | >{\centering\arraybackslash}m{0.5cm}
    | >{\centering\arraybackslash}m{0.5cm}
    | >{\centering\arraybackslash}m{0.5cm}
    | >{\centering\arraybackslash}m{0.5cm}
|}
\hline
\multirow{5}{*}{\textbf{$\beta$}}
&
\multirow{5}{*}{\rotatebox[origin=c]{90}{\textbf{level-set}}}
&
\multirow{5}{*}{%
    \rotatebox[origin=c]{90}{%
        \shortstack[c]{\textbf{interior-point}\\\textbf{level-set}
        }%
    }%
}
& \multicolumn{6}{c|}{} \\
& & & \multicolumn{6}{c|}{\textbf{DIP-set}} \\
& & & \multicolumn{6}{c|}{} \\
\cline{4-9}
& & & \multicolumn{3}{c|}{\textbf{direct}} & \multicolumn{3}{c|}{\textbf{indirect}} \\ 

\cline{4-9}
 & & & lin  & exp & log & 0.1 & 0.5 & 1.0 \\ 
\Xhline{0.8pt}

$0.125$ & 642 & 234 &     &     &     & 121 & 128 & 125 \\
$0.25$  & 534 & 154 & 148 & 181 & 163 & 109 & \cellcolor{indirect} 101 & 106 \\
$0.375$ & 498 & \cellcolor{inner}147 &     &     &     & 105 & 121 & 109 \\
$0.5$   & 478 & 149 & 152 & 146 & 167 & 108 & 113 & 119 \\
$0.625$ & 487 & 159 &     &     &     & 127 & 133 & 139 \\
$0.75$  & 456 & 184 & 133 & 130 & 135 & 149 & 161 & 169 \\
$0.875$ & 548 & 252 &     &     &     & 213 & 219 & 212 \\
\hline

\end{tabular}
};

\draw [decorate,decoration={brace,mirror,amplitude=6pt}]
  ($(tbl.south west)+(3.2cm,0)$) -- ($(tbl.south west)+(6.0cm,0)$)
  node[midway,below=6pt,align=center]{radius interpolation};

\draw [decorate,decoration={brace,mirror,amplitude=6pt}]
  ($(tbl.south west)+(6.0cm,0)$) -- ($(tbl.south west)+(8.8cm,0)$)
  node[midway,below=6pt,align=center]{convergence tolerance};

\end{tikzpicture}
\end{table}

In this initial benchmark, the indirect DIP-set regularization clearly outperforms the interior-point level-set. To assess the robustness of this observation, the next subsections extensively tests the best parameter strategies for both methods, indicated by the colors in Tab. \ref{tab:tab2}.
\subsection{Benchmark on power-sector problem}

We test the two selected parameter strategies across 17 configurations of the power-sector problem, varying the foresight, spatial scope, and SP decomposition, as shown in Tab. \ref{tab:tab3}. Runs with 33 regions and 1 or 2 months per SP exceeded the available computational resources and had to be excluded. All runs include the full 20 scenarios for different weather years. Previous research already showed that the number of scenarios has no systematic effect on the number of iterations, but only on the time required to solve all SPs \cite{Goeke2024}. All variations differ regarding the number and composition of complicating variables: The spatial scope determines the number of complicating capacity variables. The level of foresight and the further decomposition of the scenarios, corresponding to weather years, determine the complicating storage variables.

\begin{table}[ht]
\centering
\caption{Tested configurations for power-sector problem}
\label{tab:tab3}
\renewcommand{\arraystretch}{1.2}
\scriptsize
\begin{tabular}{| c | c | c !{\vrule width 0.8pt} c | c | c |}
\hline
\multirow{4}{*}{
    \rotatebox[origin=c]{90}{%
        \shortstack[c]{\textbf{foresight}}%
    }%
}
& 
\multirow{4}{*}{
    \rotatebox[origin=c]{90}{%
        \shortstack[c]{\textbf{regions}}%
    }%
}
& 
\multirow{4}{*}{
    \rotatebox[origin=c]{90}{%
        \shortstack[c]{\textbf{months}\\\textbf{per SP}}%
    }%
} & \multicolumn{3}{c|}{}  \\
& & & \multicolumn{3}{c|}{\textbf{complicating variables}} \\
& & & \multicolumn{3}{c|}{} \\
\cline{4-6}
& & & capacity & storage & total \\
\Xhline{0.8pt}

\multirow{10}{*}{perfect} & \multirow{4}{*}{5} & 12 & \multirow{4}{*}{87} & 0 & 87 \\
\cline{3-3} \cline{5-6}
 & & 6 & & 440 & 527 \\
\cline{3-3} \cline{5-6}
 & & 2 & & 1,320 & 1,407 \\
\cline{3-3} \cline{5-6}
 & & 1 & & 2,640 & 2,727 \\
\cline{2-6}

 & \multirow{4}{*}{15} & 12 & \multirow{4}{*}{264} & 0 & 264 \\
\cline{3-3} \cline{5-6}
 & & 6 & & 880 & 1144 \\
\cline{3-3} \cline{5-6}
 & & 2 & & 2,640 & 2,904 \\
\cline{3-3} \cline{5-6}
 & & 1 & & 5,280 & 5.544 \\
\cline{2-6}

 & \multirow{2}{*}{33} & 12 & \multirow{2}{*}{609} & 0 & 609 \\
\cline{3-3} \cline{5-6}
 & & 6 & & 1.960 & 2,569 \\
\hline

\multirow{7}{*}{limited} & \multirow{3}{*}{5} & 6 & \multirow{3}{*}{87} & 22 & 109 \\
\cline{3-3} \cline{5-6}
 & & 2 & & 66 & 153 \\
\cline{3-3} \cline{5-6}
 & & 1 & & 132 & 219 \\
\cline{2-6}

 & \multirow{3}{*}{15} & 6 & \multirow{3}{*}{264} & 44 & 308 \\
\cline{3-3} \cline{5-6}
 & & 2 & & 132 & 396 \\
\cline{3-3} \cline{5-6}
 & & 1 & & 264 & 528 \\
\cline{2-6}

 & 33 & 6 & 609 & 98 & 707 \\
\hline

\end{tabular}
\end{table}

\begin{figure}[!t]
\centering
\includegraphics[scale=1.0]{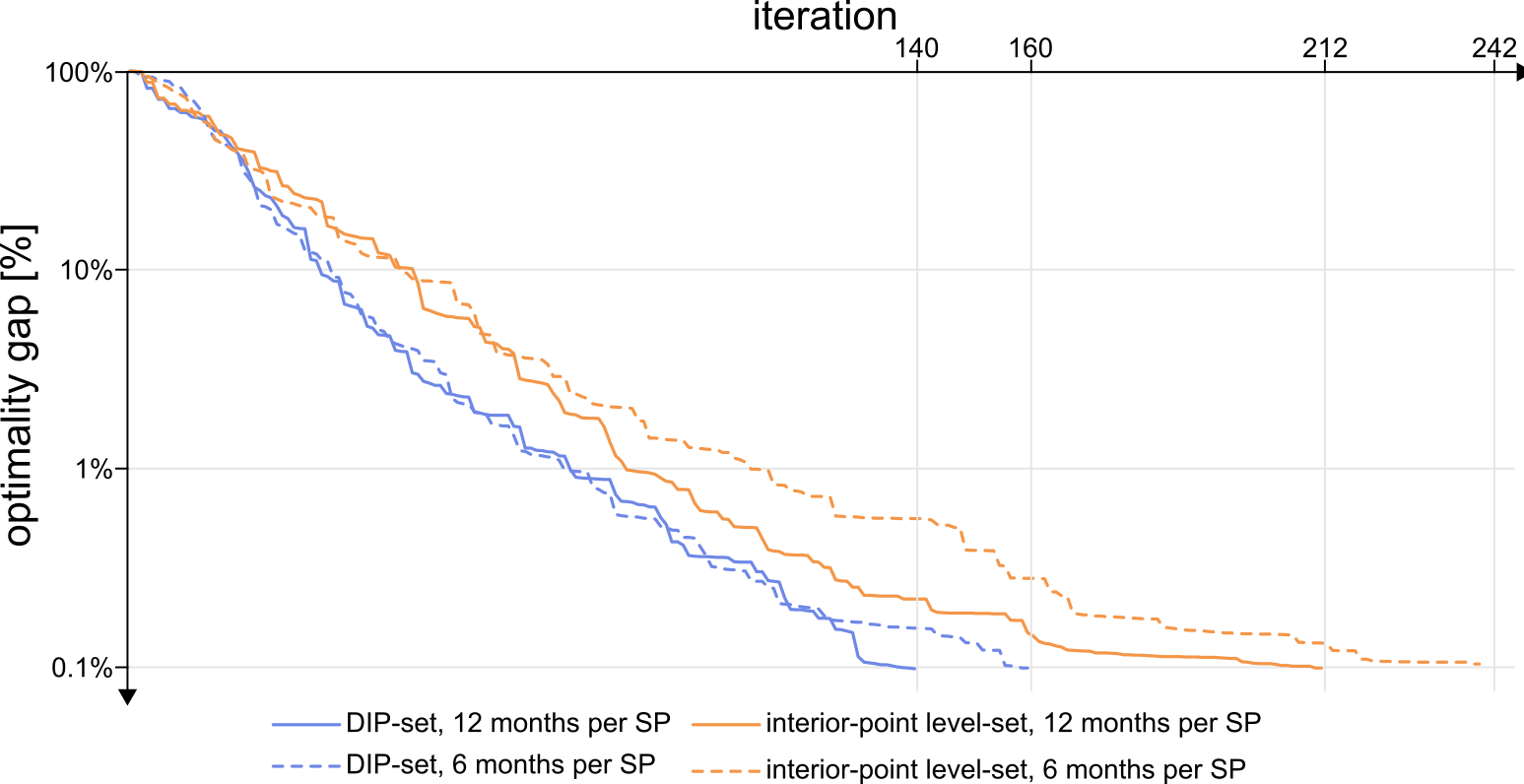}
\caption{Convergence for power-sector problem with 33 regions and perfect foresight}
\label{itrGra}
\end{figure}

Fig. \ref{itrGra} compares the two selected regularizations for the two most difficult cases: 33 regions, perfect foresight, and one year or 6 months per SP. The convergence in the logarithmic graph exhibits the tailing-off behavior typical for BD. Furthermore, DIP-set regularization outperforms the interior-point level-set by 34.0\% for 12 months per SP, i.e., in the absence of complicating storage variables, and by 33.9\% for 6 months per SP. The comparison between the 12- and 6-month per-SP also shows the expected increase in iterations with the number of complicating (storage) variables.

Tab. \ref{tab:tab4} gives a full overview of how DIP-set performed compared to inner-point level-set regularization across the 17 tested configurations. Although it is inevitable in benchmarking BD that results exhibit noise, several trends are observable.

Overall, DIP-set consistently outperforms inner-point level-set regularization, but the extent of this advantage depends significantly on the composition of complicating variables. The advantage of DIP-set clearly increases with the number of complicating capacity variables, corresponding to an increase in the number of regions. Under perfect foresight, and with 12 months per SP, the advantage increases from 15.8\% with 5 regions to 34\% with 33 regions.

On the other hand, the advantage of DIP-set decreases with the number of complicating storage variables, corresponding to a decrease in months per SP. In the case of 5 regions and one month per SP, the advantage is 2.9\% for perfect foresight, and  4.8\% for limited foresight, both within the range of expected noise. Accordingly, limited foresight, i.e., fewer complicating variables for storage, slightly increases the advantage of DIP-set regularization.

The diminishing advantage of DIP-set regularization with more complicating variables for storage reflects that the trust-region constraint, the defining feature of DIP-set regularization, excludes complicating storage variables, as discussed at the beginning of the section. As a result, DIP-set and inner-point level-set regularization become increasingly similar as the share of complicating storage variables increases, and are equivalent in the absence of complicating capacity variables.

\begin{table}[ht]
\centering
\caption{Performance of DIP-set regularization for power-sector problem}
\label{tab:tab4}
\renewcommand{\arraystretch}{1.2}
\scriptsize
\begin{tabular}{| c | c | c !{\vrule width 0.8pt} c | c | c |}
\hline
\multirow{4}{*}{
    \rotatebox[origin=c]{90}{%
        \shortstack[c]{\textbf{foresight}}%
    }%
}
& 
\multirow{4}{*}{
    \rotatebox[origin=c]{90}{%
        \shortstack[c]{\textbf{regions}}%
    }%
}
& 
\multirow{4}{*}{
    \rotatebox[origin=c]{90}{%
        \shortstack[c]{\textbf{months}\\\textbf{per SP}}%
    }%
} & \multicolumn{3}{c|}{\multirow{2}{*}{\textbf{number of iterations}}}   \\
& & & \multicolumn{3}{c|}{}  \\
\cline{4-6}
& & & \multirow{2}{*}{DIP-set} & interior-point & \multirow{2}{*}{$\Delta[\%]$} \\
& & &  & level-set &  \\
\Xhline{0.8pt}

\multirow{10}{*}{perfect} & \multirow{4}{*}{5} & 12 & 32 & 38 & -15.8 \\
\cline{3-6}
 & & 6 & 35 & 43 & -18.6 \\
\cline{3-6}
 & & 2 & 53 & 49 & -14.3 \\
\cline{3-6}
 & & 1 & 67 & 65 & -2.9 \\
\cline{2-6}

 & \multirow{4}{*}{15} & 12 & 94 & 104 & -9.6 \\
\cline{3-6}
 & & 6 & 72 & 122 & -41.0 \\
\cline{3-6}
 & & 2 & 76 & 100 & -24.0 \\
\cline{3-6}
 & & 1 & 88 & 114 & -22.8 \\
\cline{2-6}

 & \multirow{2}{*}{33} & 12 & 140 & 212 & -34.0 \\
\cline{3-6}
 & & 6 & 160 & 242 & -33.9 \\
\hline

\multirow{7}{*}{limited} & \multirow{3}{*}{5} & 6 & 34 & 42 & -19.0 \\
\cline{3-6}
 & & 2 & 33 & 40 & -17.5 \\
\cline{3-6}
 & & 1 & 36 & 37 & -4.8 \\
\cline{2-6}

 & \multirow{3}{*}{15} & 6 & 72 & 122 & -41.0 \\
\cline{3-6}
 & & 2 & 80 & 104 & -23.1 \\
\cline{3-6}
 & & 1 & 74 & 98 & -24.5 \\
\cline{2-6}

 & 33 & 6 & 138 & 270 & -48.9 \\
\hline

\end{tabular}
\end{table}

\subsection{Benchmark on energy-system problem}

To assess the consistency of the previous results for the power-system problem, we test the two selected parameter strategies across six configurations of the energy-system problem. Tab. \ref{tab:tab32} provides an overview of the tested configurations, analogous to Tab. \ref{tab:tab3}. 
\begin{table}[ht]
\centering
\caption{Tested configurations for energy-system problem}
\label{tab:tab32}
\renewcommand{\arraystretch}{1.2}
\scriptsize
\begin{tabular}{| c | c | c !{\vrule width 0.8pt} c | c | c |}
\hline
\multirow{4}{*}{
    \rotatebox[origin=c]{90}{%
        \shortstack[c]{\textbf{foresight}}%
    }%
}
& 
\multirow{4}{*}{
    \rotatebox[origin=c]{90}{%
        \shortstack[c]{\textbf{region}}%
    }%
}
& 
\multirow{4}{*}{
    \rotatebox[origin=c]{90}{%
        \shortstack[c]{\textbf{months}\\\textbf{per SP}}%
    }%
} & \multicolumn{3}{c|}{}  \\
& & & \multicolumn{3}{c|}{\textbf{complicating variables}} \\
& & & \multicolumn{3}{c|}{} \\
\cline{4-6}
& & & capacity & storage & total \\
\Xhline{0.8pt}

\multirow{6}{*}{perfect} & \multirow{2}{*}{5}  & 2 & 358 & 516 & 874 \\
\cline{3-6}
 &                     & 1 & 358 & 1,032 & 1,390 \\
\cline{2-6}
 & \multirow{2}{*}{15} & 2 & 968 & 996 & 1,964 \\
\cline{3-6}
 &                     & 1 & 968 & 1,992 & 2,960 \\
\cline{2-6}
 & \multirow{2}{*}{33} & 2 & 1,966 & 1,896 & 3,862 \\
\cline{3-6}
 &                     & 1 & 1,966 & 3,792 & 5,758 \\
\hline

\multirow{6}{*}{limited} & \multirow{2}{*}{5}  & 2 & 358 & 258 & 616 \\
\cline{3-6}
 &                     & 1 & 358 & 516 & 874 \\
\cline{2-6}
 & \multirow{2}{*}{15} & 2 & 968 & 498 & 1,466 \\
\cline{3-6}
 &                     & 1 & 968 & 996 & 1,964 \\
\cline{2-6}
 & \multirow{2}{*}{33} & 2 & 1,966 & 948 & 2,914 \\
\cline{3-6}
 &                     & 1 & 1,966 & 1,896 & 3,862 \\
\hline

\end{tabular}
\end{table}

\begin{figure}[!t]
\centering
\includegraphics[scale=1.0]{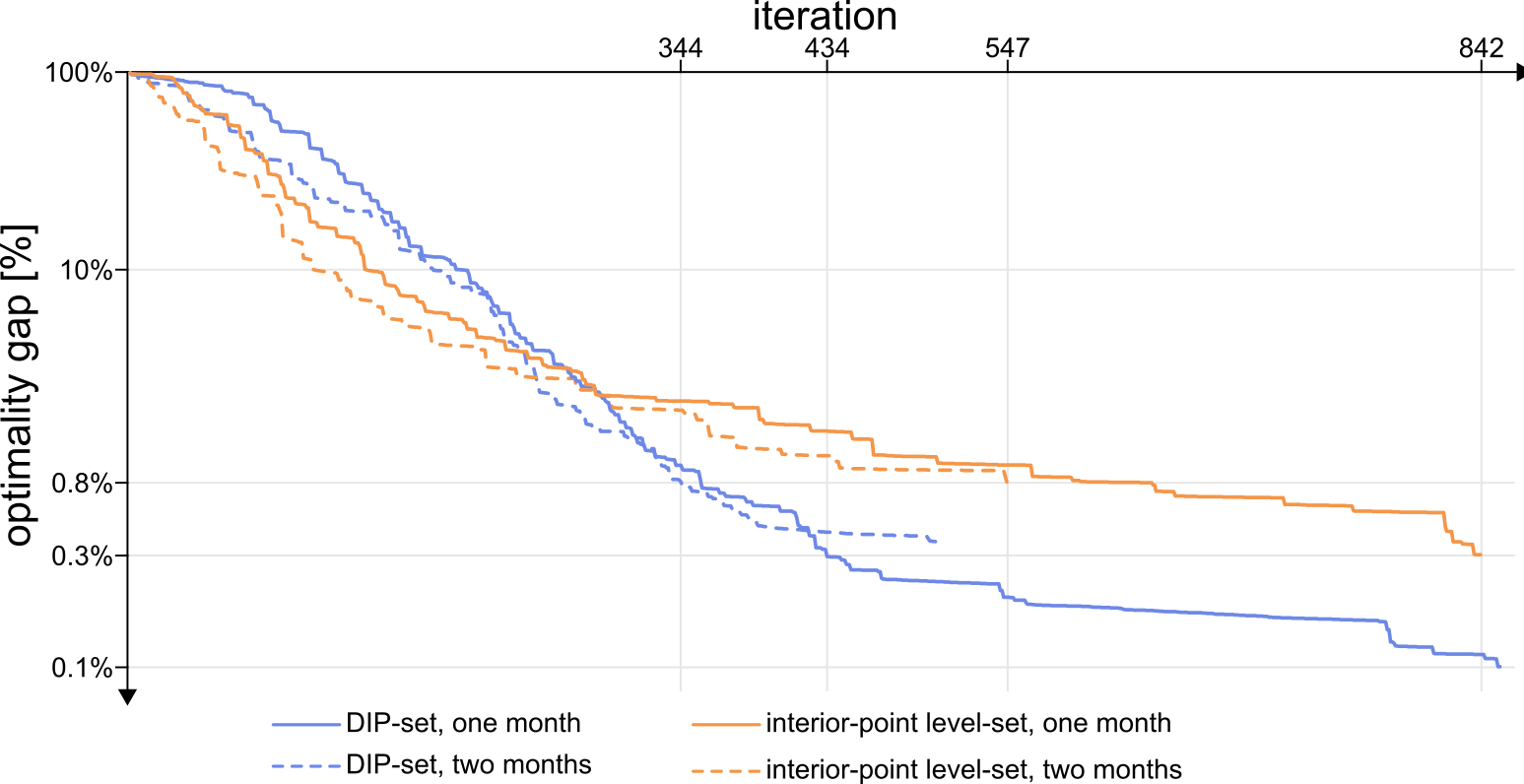}
\caption{Convergence for energy-system problem with 33 regions and two months per SP}
\label{itrGra2}
\end{figure}

Fig. \ref{itrGra2} compares the two regularizations for the 33-region case with limited foresight. Within the five-day computation time limit, only DIP-set regularization with one month per SP achieved the targeted optimality gap of 0.1\%. The interior-point regularization reached a gap of 0.34\% at the time limit; the corresponding DIP-set setup achieved this gap after only 434 iterations. With two months per SP, the DIP-set regularization shows a similar advantage.

As in the power-sector problem, the convergence exhibits strong tailing-off behavior and a substantial advantage of the DIP-set regularization. A notable difference in convergence, also observed in other runs of the energy-system problem, is that the DIP-set regularization initially converges more slowly than the interior-point level-set method. Only at an optimality gap of around 3\%, the DIP-set regularization eventually surpasses the interior-point level-set.

The trust-region constraint in the DIP-set regularization explains the initially slower convergence. Whether proximity to a reference solution enforced by the trust-region is beneficial compared to random solutions selected by the interior-point level-set depends on the quality of the reference solution. In the initial iterations, the quality of the reference solution depends on the heuristic pre-solve, and the more capacity variables the problem has, the larger the absolute error of the heuristic pre-solve and the greater the potential disadvantage of the DIP-set regularization. Once the reference solution improves in subsequent iterations, the DIP-set regularization capitalizes on the trust region's advantages and mitigates BD's tailing-off behavior.

Tab. \ref{tab:tab42} provides a full overview of the performance for the energy-system problem. The results confirm the general results and the advantage of DIP-set regularization, previously observed in the power-sector problem. The larger number of capacity variables increases the benefits of DIP-set regularization, reaching around 50\% for the largest configuration that includes all regions.

\begin{table}[ht]
\centering
\caption{Performance of DIP-set regularization for energy-system problem}
\label{tab:tab42}
\renewcommand{\arraystretch}{1.2}
\scriptsize
\begin{tabular}{| c | c | c !{\vrule width 0.8pt} c | c | c |}
\hline
\multirow{4}{*}{
    \rotatebox[origin=c]{90}{%
        \shortstack[c]{\textbf{foresight}}%
    }%
}
& 
\multirow{4}{*}{
    \rotatebox[origin=c]{90}{%
        \shortstack[c]{\textbf{region}}%
    }%
}
& 
\multirow{4}{*}{
    \rotatebox[origin=c]{90}{%
        \shortstack[c]{\textbf{months}\\\textbf{per SP}}%
    }%
} & \multicolumn{3}{c|}{}   \\
& & & \multicolumn{3}{c|}{\textbf{number of iterations}}  \\
& & & \multicolumn{3}{c|}{} \\
\cline{4-6}
& & & DIP & inner & $\Delta[\%]$ \\
\Xhline{0.8pt}

\multirow{6}{*}{perfect} & \multirow{2}{*}{5}  & 2 & 191 & 350 & -45.4 \\
\cline{3-6}
 &                     & 1 & 178 & 223 & -20.2 \\
\cline{2-6}
 & \multirow{2}{*}{15} & 2 & 536 & 1,238 & -56.7 \\
\cline{3-6}
 &                     & 1 & 455 & 840 & -45.8 \\
\cline{2-6}
 & \multirow{2}{*}{33} & 2 & 566 & 1,090 & -48.1* \\
\cline{3-6}
 &                     & 1 & 744 & 934 & -20.3* \\
\hline

\multirow{6}{*}{limited} & \multirow{2}{*}{5}  & 2 & 156 & 286 & -45.5 \\
\cline{3-6}
 &                     & 1 & 176 & 223 & -21.1 \\
\cline{2-6}
 & \multirow{2}{*}{15} & 2 & 389 & 868 & -55.2 \\
\cline{3-6}
 &                     & 1 & 353 & 633 & -44.2 \\
\cline{2-6}
 & \multirow{2}{*}{33} & 2 & 344 & 547 & -37.1* \\
\cline{3-6}
 &                     & 1 & 434 & 842 & -48.5* \\
\hline

\end{tabular}

\vspace{3mm}
{\footnotesize
\begin{tabular}{l}
*Neither regularization achieved the targeted gap of 0.1\%. The table \\ compares the iterations required to reach the gap that both models achieved.
\end{tabular}
}
\end{table}

\section{Conclusion} \label{sec:cl}

This paper introduces DIP-set regularization for solving large capacity expansion problems. The regularization combines the advantages of existing methods by traversing the interior of the solution space while remaining close to a reference solution.

We extensively benchmark the DIP-set regularization on a power-sector problem and an energy-system problem, varying problem size and the level of foresight during operation. DIP-set consistently outperforms state-of-the-art methods; its relative advantage increases with the number of capacity decisions, but not with the complicating variables for storage levels. On problems with more than 500 capacity decisions, DIP-set outperforms the next-best regularization by 30\% to 50\% thanks to mitigating the ``tailing-off'' effect in convergence. 

Future work on BD can take several directions. A potential refinement for the DIP-set regularization is to vary the enforced proximity to the current reference solution based on its quality, thereby further reducing the number of iterations. 

Far greater potential for improvement lies in machine learning-assisted BD to reduce the time per iteration \cite{Huiwen2021}. Identifying critical SPs and prioritizing them suggests a great improvement over solving all SPs in each iteration. Moreover, such priorization can serve as a starting point for asynchronous parallelization schemes to improve the current parallelization efficiency, which is only around 20\% \cite{Li2026}.

Beyond capacity expansion in energy planning, our results are generally applicable to solving large optimization problems with BD. This proposed use of BD and the recent advancements in first-order methods for large linear problems are not mutually exclusive, but complementary \cite{Applegate2025}. Since the problems suited for first-order methods are still at least one magnitude smaller than the problems solved with BD in this paper, first-order methods could replace the interior-point solvers currently solving the TP and SPs within BD.

\section*{Acknowledgments}
We thank Lukas Barner for his critical input on the methods developed in this paper and Jacob Mannhardt for his feedback on the written manuscript.

\printbibliography
%

\newpage

 


\vfill

\end{document}